\documentclass[10pt,twocolumn,letterpaper]{article}

\usepackage{iccv}
\usepackage{times}
\usepackage{epsfig}
\usepackage{graphicx}
\usepackage{amsmath}
\usepackage{amssymb}

\usepackage[pagebackref=true,breaklinks=true,letterpaper=true,colorlinks,bookmarks=false]{hyperref}

\newcommand{\NEW}[1]{{\color{black}{#1}}} 

\usepackage{ifpdf}
\usepackage{graphicx}
\ifpdf
  
\else
\fi
\usepackage{color}
\definecolor{cyan}{cmyk}{1,0,0,0}
\definecolor{darkgreen}{rgb}{0,0.5,0}
\definecolor{orange}{rgb}{1,0.5,0}
\definecolor{magenta}{cmyk}{0,1,0,0}
\definecolor{darkyellow}{cmyk}{0,0,0.75,0}

\usepackage{amssymb} 
\usepackage{amsmath} 
\usepackage{algorithm} 
\usepackage{algorithmicx}
\usepackage{algpseudocode}
\usepackage{gensymb}
\usepackage{float}
\usepackage{soul}
\usepackage{array}
\usepackage{multirow}
\usepackage{hhline}
\usepackage{setspace}
\usepackage{subfig}
\makeatletter
\renewcommand{\ALG@beginalgorithmic}{\small}
\makeatother

\newcommand{\DELETE}[1]{} 
\newcommand{\IGNORE}[1]{}
\usepackage{datenumber}
\usepackage{calc}

\newcounter{datetoday}
\newcounter{diffyears}
\newcounter{diffmonths}
\newcounter{diffdays}

\newcommand{\difftoday}[3]{%
      \setmydatenumber{datetoday}{\the\year}{\the\month}{\the\day}%
      \setmydatenumber{diffdays}{#1}{#2}{#3}%
      \addtocounter{diffdays}{-\thedatetoday}%
      \ifnum\value{diffdays}>0
        \def\diffbefore{}%
        \def\diffafter{left}%
      \else
        \def\diffbefore{}%
        \def\diffafter{ago}%
        \setcounter{diffdays}{-\value{diffdays}}%
      \fi
      \setcounter{diffyears}{\value{diffdays}/365}%
      \setcounter{diffdays}{\value{diffdays}-365*\value{diffyears}}%
      \setcounter{diffmonths}{\value{diffdays}/30}%
      \setcounter{diffdays}{\value{diffdays}-30*\value{diffmonths}}%
      \diffbefore
      \ifnum\value{diffyears}=0
      \else
        \ifnum\value{diffyears}>1
            \thediffyears\space years,
        \else
            \thediffyears\space year,
        \fi
      \fi
      \ifnum\value{diffmonths}=0
      \else
        \ifnum\value{diffmonths}>1
            \thediffmonths\space months
        \else
            \thediffmonths\space month
        \fi
      \fi
      \ifnum\value{diffdays}=0
      \else
        \ifnum\value{diffdays}>1
            \thediffdays\space days
        \else
            \thediffdays\space day
        \fi
      \fi
      \diffafter
}

\usepackage{url}

\usepackage{breakurl}
	
 \iccvfinalcopy 

\ificcvfinal\fi
\begin{document}

\title{
HDR Video Reconstruction with Tri-Exposure Quad-Bayer Sensors
}
 
\author{Yitong Jiang  ~ ~ ~ Inchang Choi ~~~ Jun Jiang  ~ ~ ~ Jinwei Gu\\
SenseBrain\\
{\tt\small \{jiangyitong,inchangchoi,jiangjun,gujinwei\}@sensebrain.ai
}
}



\twocolumn[{%
\renewcommand\twocolumn[1][]{#1}%
\maketitle
\begin{center}
    \centering
    \captionsetup{type=figure}
    \includegraphics[width=0.96\textwidth]{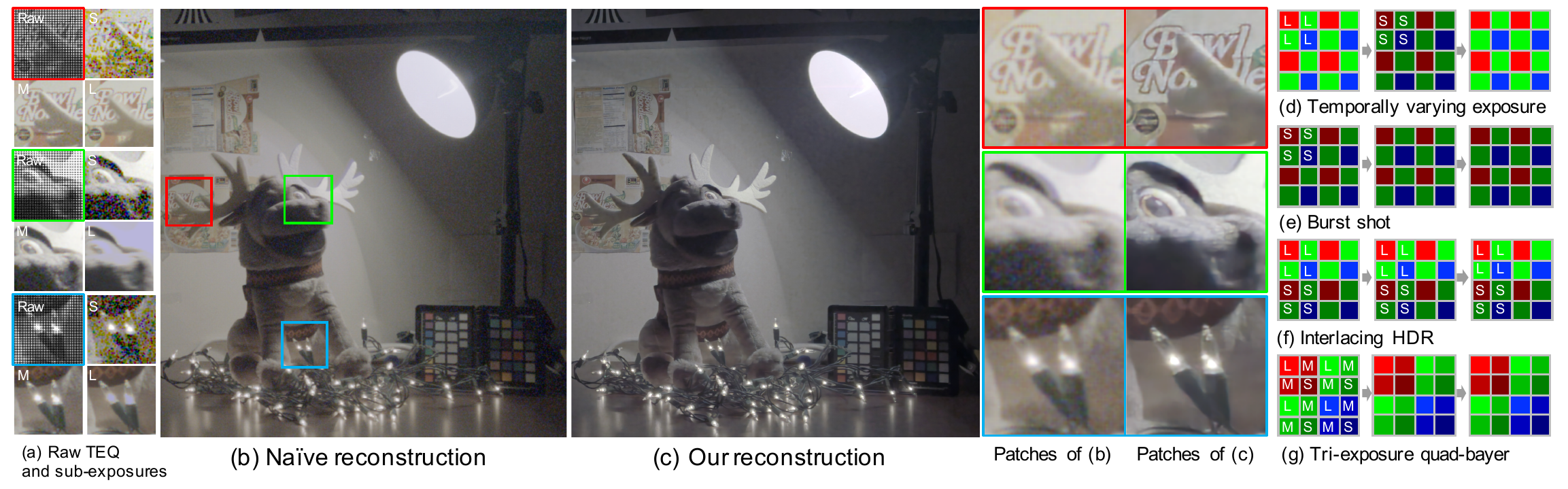}
    \vspace{-3.5mm}
    \small{\captionof{figure}{Our HDR reconstruction of a challenging scene is shown in (c). The real tri-exposure quad-bayer (TEQ) input of the scene, described in (g), and its sub-exposure images are shown in (a)\protect\footnotemark. Saturation, noise, spatial artifact, and motion blur in the sub-exposure images are attributed to the quality degradation of an engineered-interpolation-based naive reconstruction in (b). From (d) to (g), four exposure strategies for the HDR video are described. 
}\label{fig:intro}}
\end{center}%
}]
\footnotetext{L, M, and S on the color filter stand for long, middle, and short exposures, respectively.}


\begin{abstract}

We propose a novel high dynamic range (HDR) video reconstruction method with new tri-exposure quad-bayer sensors. Thanks to the larger number of exposure sets and their spatially uniform deployment over a frame, they are more robust to noise and spatial artifacts than previous spatially varying exposure (SVE) HDR video methods. Nonetheless, the motion blur from longer exposures, the noise from short exposures, and inherent spatial artifacts of the SVE methods  remain huge obstacles. Additionally, temporal coherence must be taken into account for the stability of video reconstruction. To tackle these challenges, we introduce a novel network architecture that divides-and-conquers these problems. In order to better adapt the network to the large dynamic range, we also propose LDR-reconstruction loss that takes equal contributions from both the highlighted and the shaded pixels of HDR frames. Through a series of comparisons and ablation studies, we show that the tri-exposure quad-bayer with our solution is more optimal to capture than previous reconstruction methods, particularly for the scenes with larger dynamic range and objects with motion.
\end{abstract}

\vspace{-8.0mm}
\section{Introduction}
\label{sec:intro}
\graphicspath{ {./figures/} }


Digital image sensors have a limited dynamic range (e.g., ~$60$ dB for mobile phone cameras), which is determined by the full-well capacity, dark current, and read noise. The constrained dynamic range often gives us unsatisfying portrait photographs with either excessively dark and noisy faces or completely saturated backgrounds, and it is one of the major issues that make photographing less enjoyable. To mitigate the limit, \emph{high dynamic range} (HDR) imaging was introduced~\cite{debevec1997expbracketing,gupta2013fibnoaccibracketing} and has been under significant attention for a couple of decades. In addition, \emph{HDR video}~\cite{kang2003hdrvideo,kalantari2019deephdrv,choi2017interlacing} is getting closer to our real life. It is not difficult to find TVs that support HDR video, and video sharing platforms also have started to stream HDR contents.

\begin{figure}[t]
\begin{center}
\includegraphics[width=0.99\linewidth]{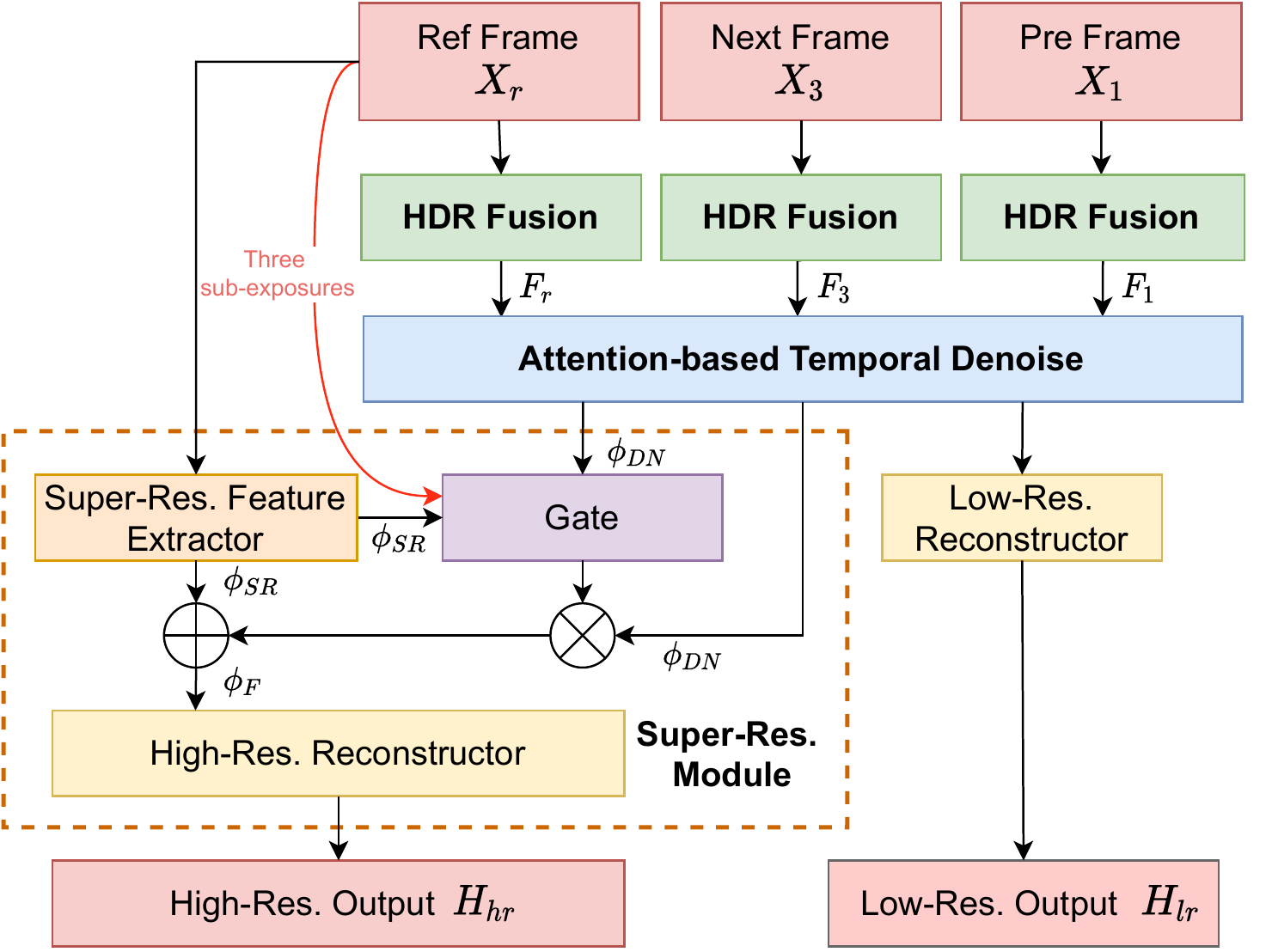}
\end{center}
\vspace{-6.0mm}
\caption{The overview of our HDR reconstruction network architecture.}
\label{fig:arch:overview}
\vspace{-3.0mm}
\end{figure}

For the acquisition of HDR video, there have been three major approaches. \emph{Temporally varying exposure} (TVE), also known as \emph{exposure bracketing}, takes multiple shots of a scene with different exposure settings as shown in Figure~\ref{fig:intro}(d), and fuses one HDR frame by weighted-averaging well-exposed pixels from adjacent frames. Unfortunately, it suffers from \emph{ghosting artifacts}~\cite{mangiatANDgibson2010hdrvideo,gryaditskaya2015motionawarehdrv} for dynamic objects, and shows an intrinsic trade-off between motion and dynamic range. This could be alleviated by a \emph{burst shot} (BS)~\cite{hasinoff2016burst,liba2019nightsight,liu2014fastburstdenoising} approach, described in Figure~\ref{fig:intro}(e), that takes a considerably short exposure to minimize the motion between the frames. However, it undergoes severe quantization and noise in the dark region of scenes and requires expensive denoising algorithms~\cite{tassano2020fastdvdnet,jiang2019seemovinginthedark}. Finally, there are \emph{spatially varying exposure} (SVE) methods that are the least hindered by the ghosting artifacts and the quantization problem. In modern image sensors that support SVE, a number of the sets of pixels within one frame can take different exposures. But their HDR reconstructions have problems of reduced resolution and interlacing artifacts~\cite{nayar2000spatiallyvarying,gupta2013fibnoaccibracketing,cho2014singleshothdr, cogalan2020interlacing}.

We propose a novel HDR video reconstruction method for new SVE sensors: \emph{tri-exposure quad-bayer sensors}~\cite{samsung20tetracell,quadbayer_explained19,redmi19}. As shown in Figure~\ref{fig:intro}(g), the quad-bayer sensors spatially extend each color filter of the bayer to four neighboring pixels. Their tri-exposure mode enables to set three different exposure settings within each color. Comparing to the interlacing HDR methods~\cite{cogalan2020interlacing, choi2017interlacing} shown in (f), which has represented the SVEs, it samples more exposure sets uniformly over the image. Therefore it is more robust to the noise and the spatial artifacts. Nonetheless, a naive HDR video reconstruction\footnote{A traditional reconstruction method using engineered interpolations.} method (b) would suffer from three problems described in (a): the motion blur from longer exposures, the noise from short exposures, and inherent spatial artifact and resolution degradation. In addition, the temporal coherence must be taken into account for the better stability of the video.

To tackle the problem, we introduce novel network architecture, shown in Figure~\ref{fig:arch:overview}, that divides-and-conquers three problems. It is modularized by the \emph{HDR feature fusion} module that performs HDR fusion in the feature space to address the motion blur, the \emph{attention-based temporal denoising} module that performs the multi-frame noise reduction and maintains the temporal coherence, and the \emph{super-resolution} module that alleviates the remaining spatial artifact and resolution problems as shown in Figure~\ref{fig:intro}(c). To better adapt the network to the large dynamic range, we also propose \emph{LDR-reconstruction loss} that takes equal contributions from highlighted and shaded pixels in output HDR frames. We provide a thorough comparison with other HDR video methods, including TVEs, BSs, and SVEs, and ablation studies. They show that the tri-exposure quad-bayer with our proposed solution is more optimal to capture HDR video than the previous HDR reconstruction methods, particularly for the scenes with larger dynamic range and dynamic objects. All the code and the data will be available upon publication.

\section{Related Work}
\label{sec:relatedwork}
\vspace{-1mm}
\paragraph{HDR using Temporally Varying Exposure}
An exposure strategy of alternating different exposures called \emph{exposure bracketing} was introduced by~\cite{debevec1997expbracketing,gupta2013fibnoaccibracketing} to capture HDR images. Kang et al.~\cite{kang2003hdrvideo} extended it to video by utilizing optical flow to align neighborhood frames to a reference frame. In the succeeding research, more robust motion estimation methods relying on block-based motion~\cite{mangiatANDgibson2010hdrvideo, mangiatANDgibson2011hdrvideoII} and patch-based synthesis~\cite{kalantari2013patchhdrv} were proposed. Subsequently, Gryaditskaya et al.~\cite{gryaditskaya2015motionawarehdrv} performed motion-aware exposure bracketing by considering the perceptual importance of motion and dynamic range, and Kalantari et al.~\cite{kalantari2019deephdrv} presented a CNN-based HDR video reconstruction. Nonetheless, none of these methods were completely free from the ghosting artifact caused by fast large motion and the extensive loss of rigidity. In contrast, our HDR video reconstruction is more robust to motion for the inherent robustness of the SVEs and our attention-based temporal denoising that does not rely on explicit motion estimation.

\begin{figure*}[t]
\centering
\includegraphics[width=0.99\linewidth]{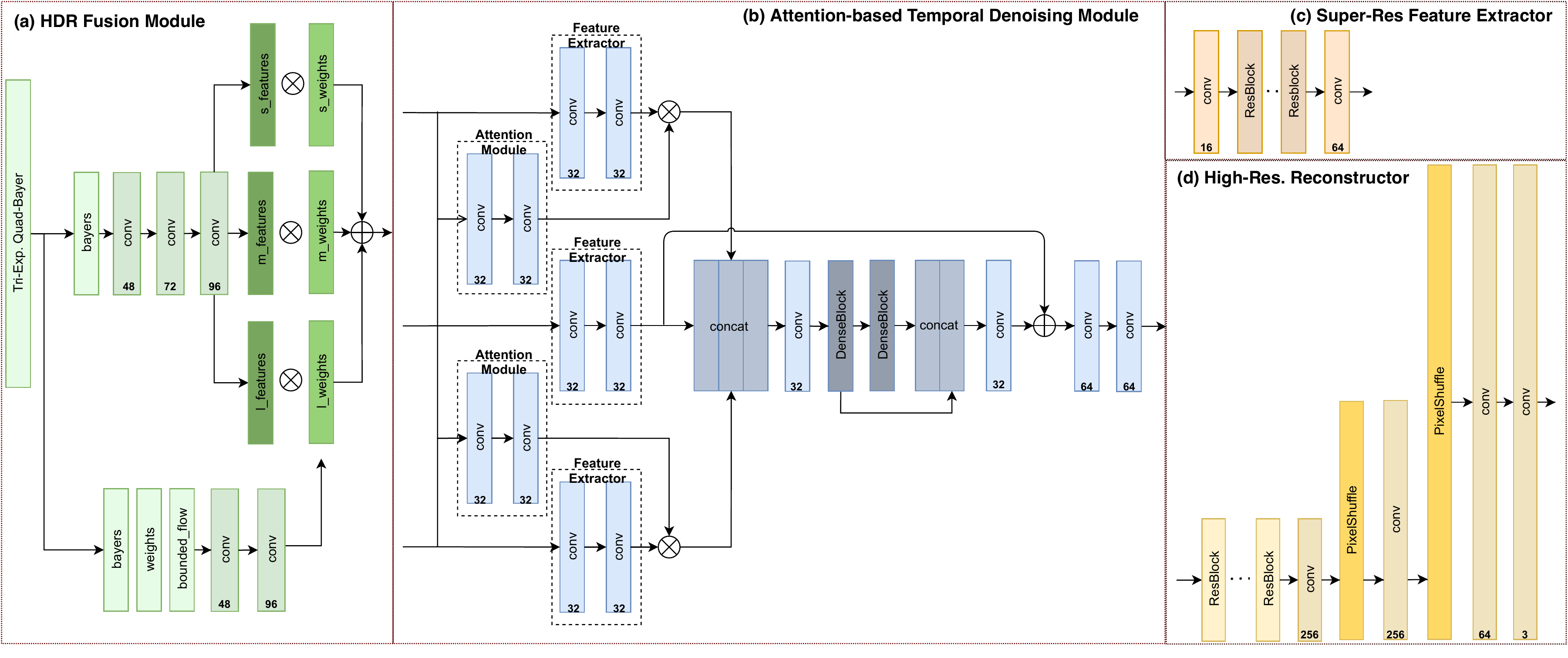}
\vspace{-4.0mm}
\caption{The modules in our reconstruction network are described here. The HDR fusion module is shown in (a). The attention-based temporal denoising module is depicted in (b). The super-resolution feature extractor and the high-resolution reconstructor are described in (c) and (d), respectively.}
\vspace{-5.5mm}
\label{fig:arch:detail}
\end{figure*}

\paragraph{HDR from Burst Shots}
With recent emerging interests in low-light imaging~\cite{chen18seeinthedark,dong11lowlight,malm2007lowlight}, the burst-shot-based HDR image algorithms have obtained great attention~\cite{liu2014fastburstdenoising,hasinoff2016burst,liba2019nightsight}. It takes multiple shots of images with the same short exposure. The fixed exposure makes the motion estimation more robust 	since the intensity level and the noise level do not change over the burst sequence. Hasinoff et al.~\cite{hasinoff2016burst} proposed an efficient system for the burst-shot-based HDR, and Liba et al.~\cite{liba2019nightsight} improved it by introducing a superior exposure scheduler and a new merging algorithm. Extending the burst imaging algorithm to video is straightforward, but it suffers severe noise and quantization problems for the dark region in the video. Therefore, strong and computationally demanding video denoising algorithms~\cite{maggioni2011vbm4d,godard2018deepburstdenoising,jiang2019seemovinginthedark,tassano2019dvdnet,tassano2020fastdvdnet} must be accompanied as post-processing. On the other hand, the HDR video from our tri-exposure quad-bayer alleviates the noise and the quantization by taking short, middle, and long exposures together in each frame while maintaining the robustness to the ghosting artifact.

\paragraph{HDR using Spatially Varying Exposure}
The other HDR image/video acquisition is to utilize the spatially varying exposure that takes multiple different exposures within a single frame~\cite{nayar2000spatiallyvarying, gu2010codedrollingshutter}. Although this approach is less suffered from the ghosting artifacts, it suffers from spatial artifacts on under-/over-exposed pixels and high contrast regions. Heide et al.~\cite{heide2014flexisp} and Cho et al.~\cite{cho2014singleshothdr} addressed it by global optimization with image priors, and an adaptive filter was used in~\cite{hajisharif2014alternatinggain}. Serrano et al.~\cite{serrano2016csc} proposed to use learned image priors using a convolutional sparse coding model. Choi et al.~\cite{choi2017interlacing} and çoğalan and Akyüz~\cite{cogalan2020interlacing} introduced an HDR video reconstruction algorithm for interlacing sensors with two exposure sets using joint sparse coding and deep learning, respectively. But they were vulnerable to artifacts and resolution degeneration in the vertical direction. Conversely, three exposure sets of the tri-exposure quad-bayer, which are uniformly distributed over the image, and our novel reconstruction network produce better HDR video with less noise and spatial artifacts.

\paragraph{Learning-based HDR Reconstruction}
Recently, a series of works that utilizes deep convolutional neural networks to reconstruct HDR images from the temporally varying exposure has been presented. Their focus was mainly on solving the ghosting artifacts. Kalantari et al.~\cite{kalantari2017hdri} proposed to use learned optical flow for the alignment, and it  was extended to video in~\cite{kalantari2019deephdrv}. Larger network architecture and a spatial attention module mitigated the motion problem by Wu et al.~\cite{wu2018hdri} and Yan et al.~\cite{yan2019attention}. For the spatially varying exposure, An et al.~\cite{an2017deepinterlaicng} and Coğalan and Akyüz~\cite{cogalan2020interlacing} proposed a CNN-based reconstruction method for an interlacing sensor. All of these methods apply a simple global tone mapping with a fixed parameter ~\cite{kalantari2017hdri,kalantari2019deephdrv,wu2018hdri,yan2019attention} on output and label HDR images to make the network adapt to a large dynamic range. This not optimal for preserving the local contrast and removing the noise in the dark, since relatively dark pixels cannot contribute enough in the computation of the loss. In contrast, our LDR reconstruction loss that uses simulated LDR images in the loss function leads to superior HDR video reconstruction.

\section{Our Method}
\label{sec:ourmethod}
\graphicspath{ {./figures/} }

As shown in Figure~\ref{fig:arch:overview}, our HDR reconstruction network takes three consecutive raw frames of tri-exposure quad-bayers $\{X_1, X_2, X_3\}$ as inputs and generates a clean HDR frame $\widetilde{H}$. To clarify, we denote $X_2$ as $X_r$ to indicate the reference frame and keep the same convention for other variables as well. The network consists of three modules. First, the \emph{HDR fusion} module produces HDR features by fusing the sub-exposures\footnote{Three sub-sampled images with the different exposures. They are one-fourth of the input raw in size since the sub-sampling is with respect to the exposure and the color.} in each tri-exposure quad-bayer $X_i$. Second, the \emph{attention-based temporal denoising} module takes the HDR features of the three input frames to remove the noise in the feature space. Note that the outputs of the HDR fusion module and the denoising module have the half resolution of the raw inputs. Finally, the \emph{super-resolution} module removes the spatial artifacts and recovers the original resolution. This module utilizes the high-resolution feature extracted from the raw input $X_r$. It is merged with the denoised low-resolution by a learned gate operation~\cite{zhang2018gatedfusion}, and the high-resolution reconstructor generates a final HDR frame. The following subsections describe the details of each module. 



\subsection{HDR Fusion Module}
\label{sec:ourmethod:hdrfusion}
This module performs the HDR fusion of three different exposures in a tri-exposure quad-bayer in the feature space. As shown in Figure~\ref{fig:arch:detail}(a), one branch of the module produces features from three exposures: $f_S$, $f_M$, and $f_L$. In the other branch, the weights, $w_S$, $w_M$, and $w_L$ are estimated. This estimation includes the conventional trapezoidal intensity weight map~\cite{debevec1997expbracketing}, which is widely used in the HDR image reconstruction, and the bounded flow map~\cite{liba2019nightsight} as inputs. Having the bounded flow helps the module learn to reject the motion blur from the long exposures. The HDR feature $F$ is the weighted sum of each exposure feature.
\begin{equation}%
	F = \sum_{i} f_{i} \circ w_{i}
\end{equation}
where  $\circ$ denotes the Hadamard product, and $i$ is a variable for indicating one of three exposures $\{S, M, L\}$. Here, we produce three HDR features $\{F_1, F_r, F_3\}$ for $\{X_1, X_r, X_3\}$ using a shared HDR fusion module.

\subsection{Attention-based Temporal Denoising Module}
\label{sec:ourmethod:temporaldenoising}
Our temporal denoising utilizes attention modules~\cite{yan2019attention,tassano2020fastdvdnet} to address the misalignment between the frames instead of explicitly estimating the motion. The attention modules extract useful features from different frames to refine the reference frame. As shown in Figure~\ref{fig:arch:detail}(b), we compute the attention $A_j$ on $F_j$ with respect to the reference HDR feature $F_r$ as follows:

\begin{equation}%
	A_{j} = a(F_{j},F_{r}),
\end{equation}
where $a\left(\cdot\right)$ is the attention module that consists of two convolutional layers and $j \in \{1, 3\}$. Then the attention $A_j$ is used to attend the refined non-reference HDR feature $\bar{F}_{j}$ that is produced from $F_j$ by the feature extractor. 

\begin{equation}%
	Z_{j} = A_j \circ \bar{F}_{j}
\end{equation}
Note that the attention modules share the weight, and so do the feature extractor.
Then, we stack the attended feature $Z_1$ and $Z_3$ with the refined reference HDR feature $\bar{F}_{r}$.
\begin{equation}%
	Z_{t} = concat(Z_{1},\bar{F}_{r},Z_{3})
\end{equation}
$Z_{t}$ is passed through a convolutional layer followed by dilated residual dense blocks~\cite{yan2019attention}. After one more convolutional layer, it goes through a skip connection that adds $\bar{F}_{r}$ and two more convolutional layers to make the denoised HDR feature $\phi_{DN}$. 

As shown in Figure~\ref{fig:arch:overview},  the temporal denoising module engages with the low-resolution reconstructor. The low-resolution reconstructor consists of three convolutional layers that reconstruct a low-resolution HDR frame with the quarter size of  the raw input. As described in Section~\ref{ourmethod:trainingloss}, we compute a loss on this low-resolution reconstruction to make the denoising module focus on learning its own task.
\subsection{Super-Resolution Module}
The super-resolution module consists of the super-resolution feature extractor and the high-resolution reconstructor. In the super-resolution feature extractor, shown in Figure~\ref{fig:arch:overview}(c), we use eight ResBlocks to extract the super-resolution feature  $\phi_{SR}$ from an input raw quad-bayer. This is intended to fully utilize spatial information. The gate module performs a feature fusion to form an input to the high-resolution reconstructor, which is as shown in Figure~\ref{fig:arch:detail}(d):
\begin{equation}%
\phi_{F} = G_{gate}\left(\phi_{SR}, \phi_{DN}, X_r\right)\circ \phi_{DN} + \phi_{SR}.
\label{eq:tonemapping}
\end{equation}
$\phi_{F}$ is fed into the high-resolution reconstructor. First, it passes through eight ResBlocks followed by a convolutional layer. Then, two pixel-shuffle-layers are deployed to upsample the features. Each pixel-shuffle is followed by a convolutional layer. Finally, the last convolutional layer reconstructs a high-resolution HDR image with RGB colors.

\begin{figure*}[t]
\centering
\includegraphics[width=0.99\linewidth]{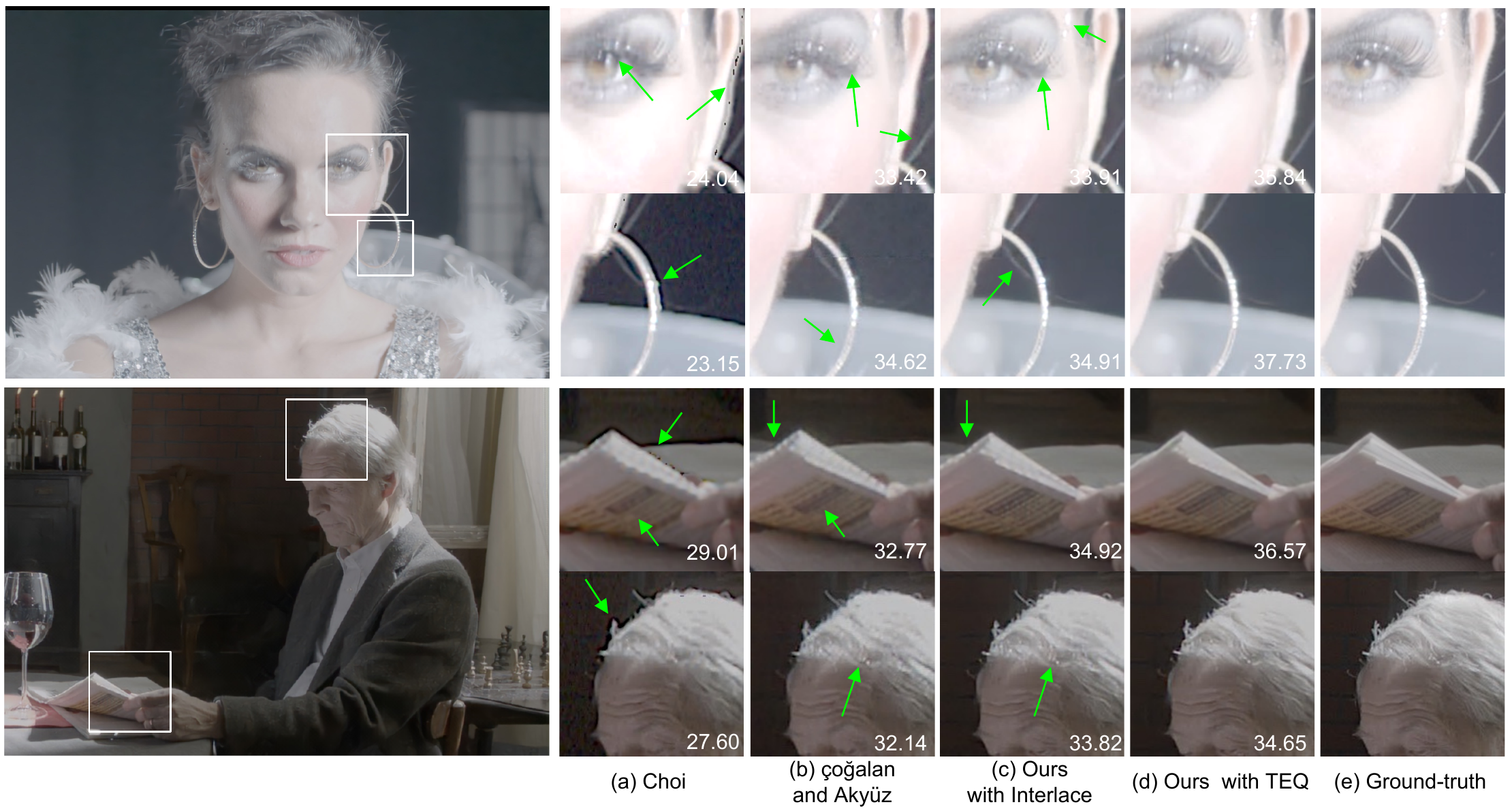}
\vspace{-4.0mm}
\caption{\NEW{We compare the proposed method to state-of-the-art interlacing HDR video methods. The methods of Choi et al.~\cite{choi2017interlacing} and çoğalan and Akyüz~\cite{cogalan2020interlacing} are shown in (a) and (b). Our proposed model trained on interlaced inputs in (c) better removes interlace artifacts. Our results from the tri-exposure quad-bayer inputs are demonstrated in (d), and (e) is the ground-truth. The PSNR values are located in each patch.}
}
\label{result:comp_interlace}
\vspace{-5.0mm}
\end{figure*}

\subsection{Training Loss}
\label{ourmethod:trainingloss}
In the previous learning-based HDR image/video reconstructions~\cite{kalantari2019deephdrv,wu2018hdri,yan2019attention}, it has been prevalent to perform global tone mapping before computing the loss function. Because of the wide intensity range of the HDR images, the loss for the dark pixels has less effect on the total loss value, and it was not able to reconstruct the dark part of the images without boosting the dark by the tone mapping. The following global tone mapping function, called $\mu$-law, was used in the previous methods:
\begin{equation}%
	T(H) = \frac{\log{(1+\mu H)}}{\log{(1+\mu)} }, 
\label{eq:tonemapping}
\end{equation}
where $H$ is a reconstructed HDR image, and $\mu$ is a parameter adjusting the amount of the dynamic range compression. 

However, for the scenes with an extensively large dynamic range, no matter how large $\mu$\footnote{The fixed value ($\mu = 5000$) has been widely adopted.} we set, the contribution of the dark pixels in the total loss is not amplified significantly enough. This will cause artifacts in the dark region. To address this, we propose a novel \emph{LDR-reconstruction loss}. For the LDR-reconstruction loss, we first simulate the LDR images from the reconstructed HDR image:

\begin{equation}%
	I^{LDR}_{i}(H) = (H\cdot t_i \cdot g_{i})^{\frac{1}{2.2}},
\end{equation}
where $t_{i}$ and $g_{i}$ is the exposure time and the gain\footnote{$t_{i}$ and $g_{i}$ are known from the moment of the image capture}, and $i$ is a variable for indicating one of three exposures $\{S, M, L\}$ in the tri-exposure quad-bayer. Then, the LDR-reconstruction loss $\mathcal{L}^{LDR}(\widetilde{H},H)$ is defined as:

\begin{equation}%
\sum_{i} \mathcal{L}(\,\omega \circ {I}^{LDR}_{i}(\widetilde{H}),\;\omega \circ I^{LDR}_{i}(H)\,),
\end{equation}
where $\widetilde{H}$ is the ground-truth HDR image, and $\omega$ is a mask indicating well-exposed pixels in the simulated LDR image. $\mathcal{L}$ could be any loss function that measures the difference of images, but we use the $\ell^1$ loss and the perceptual loss~\cite{johnson16perception}. 

Our network produces a high-resolution HDR image $H_{hr}$ as well as a low-resolution HDR image $H_{lr}$ as mentioned in Section~\ref{sec:ourmethod:temporaldenoising}. By computing the loss on the low-resolution HDR image, we enforce the temporal denoising module to focus on denoising tasks, and we achieve better HDR reconstruction as shown in Section~\ref{ablation:loss}. Our final loss function is formulated as follows:
\begin{equation}%
\mathcal{L}^{LDR}(\widetilde{H}_{hr},{H}_{hr}) + \alpha \mathcal{L}^{LDR}(\widetilde{H}_{lr},{H}_{lr}),
\label{eq:ldrreconloss}
\end{equation}
where $\widetilde{H}_{hr}$ and $\widetilde{H}_{lr}$ are the corresponding ground-truth images, and $\alpha$ is set to $0.6$.

\subsection{HDR Dataset and Raw Simulation}
We simulated tri-exposure quad-bayers from 28 HDR footages from three public datasets: Froehlich et al.~\cite{froehlich14hdrvideo}, Kronander et al.~\cite{kronander13hdrvideo}, and Azimi et al.~\cite{azimi18hdrvideo}.
From the selected HDR footages, we generate three LDR images with short, middle, and long exposure time. The ratio between adjacent exposures is set to 4.  Then, we merge the three LDR images to be a tri-exposure quad-bayer image. 

We added Gaussian noise to the LDR images. The standard deviation was randomly selected from between $4\times 10^{-3}$ to $1.6\times 10^{-2}$ for the short exposure time LDR image. The standard deviations for the middle and the long exposure are computed by multiplying the exposure ratio on it. In addition to the noise, we also simulated motion blur in the LDR images using Super Slomo~\cite{jiang18superslomo}.

\subsection{Implementation Details}
Our model is implemented in PyTorch~\cite{paszke2019pytorch}. For training, we used Adam optimizer and set the batch size and learning rate as $12$ and $1 \times 10^{-4}$, respectively. We break down the training images into the overlapping patches of $256 \times 256$ with a stride of $120$ pixels. The network was initialized by Xavier initialization~\cite{glorot10understanding}. Our method takes 0.008 seconds  to generate an HDR image from a $1920\times 1080$ quad-bayer raw with NVIDIA 2070 Super GPU. 

\section{Results}
\label{sec:results}
\graphicspath{ {./figures/} }

\begin{figure*}[t]
\centering
\includegraphics[width=0.99\linewidth]{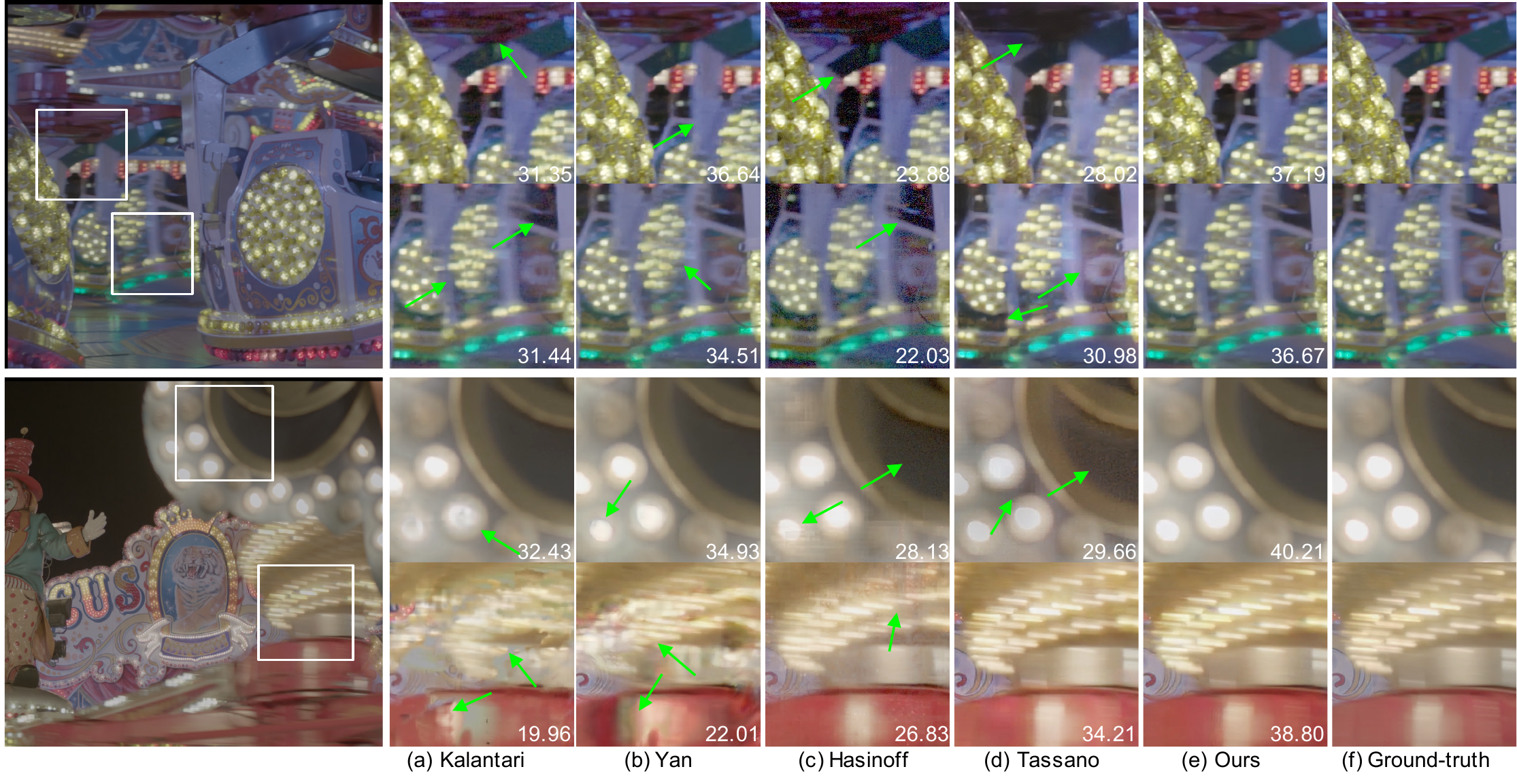}
\vspace{-4.0mm}
\caption{We compare the proposed method to state-of-the-art HDR video methods. The temporally varying exposure methods of Kalantari et al.~\cite{kalantari2019deephdrv} and Yan et al.~\cite{yan2019attention} are shown in (a) and (b). The burst shot methods of Hasinoff et al.~\cite{hasinoff2016burst} and Tassano et al.~\cite{tassano2020fastdvdnet} are shown in (c) and (d). Our results are demonstrated in (e), and (f) is the ground-truth. The PSNR values are located in each patch.
}
\label{result:comp_state_of_the_art}
\vspace{-3.0mm}
\end{figure*}

\subsection{Comparison to Interlacing HDR}
\NEW{
We conducted a comparison between our reconstruction with the TEQ and the state-of-the-art interlacing HDR video reconstruction method. In this comparison, we included a model of our proposed architecture trained with simulated interlacing bayers to verify the effectiveness of it. Figure~\ref{result:comp_interlace} shows the results of the comparison. The joint sparse coding method of Choi et al.~\cite{choi2017interlacing} in (a) suffers from noise and spatial interlacing artifacts on the over-exposed and under-exposed region. The CNN-based method of çoğalan and Akyüz~\cite{cogalan2020interlacing} in (b) better removes the noise, but it cannot recover the detail of textures. In contrast, our model trained with interlacing inputs in (c) shows better reconstruction than the previous two methods. However, our reconstruction using TEQ in (d) is more close to the ground-truth in (e). This is attributed to the uniform deployment of the exposure samples and the extra medium exposure samples in TEQ as shown in Figure~\ref{fig:intro}.

We conducted the qualitative evaluations on twelve test scenes. The averaged quality evaluations are shown in Table~\ref{table:comp_state_of_the_art_interlace}. Our method produces the best scores of PSNR-$\mu$, PuPSNR, HDR-VDP~\cite{mantiuk11hdrvdp}, and HDR-VQM~\cite{narwaria15hdrvqm}. Note that HDR-VQM is a metric to assess the HDR video quality including the notion of temporal coherence.
}
\begin{table}[t]
\centering
\resizebox{82mm}{!}{
\begin{tabular}{c | c c c c }
 \hline
 & PSNR-$\mu$ & PuPSNR & HDR-VDP & HDR-VQM  \\

 \hline\hline
Choi  & $28.32$ & $36.16$  & $40.21$ & $0.7703$\\

çoğalan and Akyüz & {$34.71$} & {$40.91$}  & {$44.69$} & {$0.8802$} \\

Ours (interlace)  & \textcolor{blue}{$35.90$} &  \textcolor{blue}{$41.56$} & \textcolor{blue}{$45.02$} & \textcolor{blue}{$0.9194$} \\

Ours (TEQ) & \textcolor{red}{$36.06$} &  \textcolor{red}{$41.64$} &  \textcolor{red}{$46.16$} & \textcolor{red}{$0.9275$}  \\

 \hline
\end{tabular}}
\caption{This table shows the quantitative evaluation of state-of-the-art HDR video methods with interlacing inputs.}
\label{table:comp_state_of_the_art_interlace}
\vspace{-5mm}
\end{table}

\begin{table}[t]
\centering
\resizebox{82mm}{!}{
\begin{tabular}{c | c c c c }
 \hline
 & PSNR-$\mu$ & PuPSNR & HDR-VDP & HDR-VQM  \\

 \hline\hline
Kalantari  & $33.27$ & $38.48$  & $44.79$ & $0.8632$\\

Yan  & \textcolor{blue}{$35.67$} &  \textcolor{blue}{$40.75$}  & \textcolor{blue}{$45.01$} & \textcolor{blue}{$0.9226$} \\

Hasinoff  & $23.73$ &  $31.62$ & $39.42$ & $0.7645$ \\

Tassano & $26.15$ & $34.10$ & $41.45$ & $0.8096$\\

Ours & \textcolor{red}{$36.06$} &  \textcolor{red}{$41.64$} &  \textcolor{red}{$46.16$} & \textcolor{red}{$0.9275$}  \\

 \hline
\end{tabular}}
\vspace{-1mm}

\caption{This table shows the quantitative evaluation of state-of-the-art HDR video methods.}
\label{table:comp_state_of_the_art}
\vspace{-3mm}
\end{table}

\subsection{Comparison to TVEs and BS}
We compared our TEQ HDR video reconstruction to previous state-of-the-art methods on challenging HDR video scenes. Specifically, the comparison includes two temporally varying exposure (TVE) HDR video methods~\cite{kalantari2019deephdrv,yan2019attention} and two burst shot HDR video methods~\cite{hasinoff2016burst,tassano2020fastdvdnet}. The same dataset and exposure settings are used for simulating the inputs with the different exposure strategies.

\begin{figure*}[t]
\centering
\includegraphics[width=0.99\linewidth]{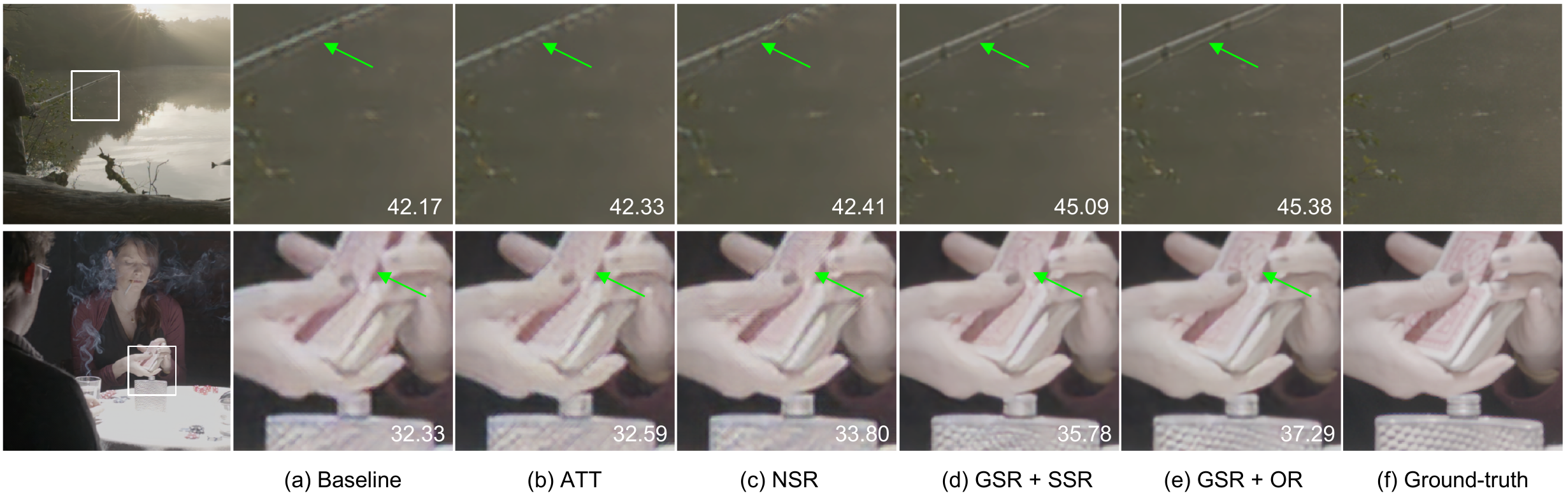}
\vspace{-1.0mm}
\caption{We investigated the effect of each module in the proposed network model. The network, GSR + OR, that has all of the proposed modules shows the best quality. Refer to Section~\ref{ablation:architecture} for the details of each network.}
\label{result:ablataion1}
\vspace{-2.0mm}
\end{figure*}

\begin{figure}[t]
\centering
\includegraphics[width=0.99\linewidth]{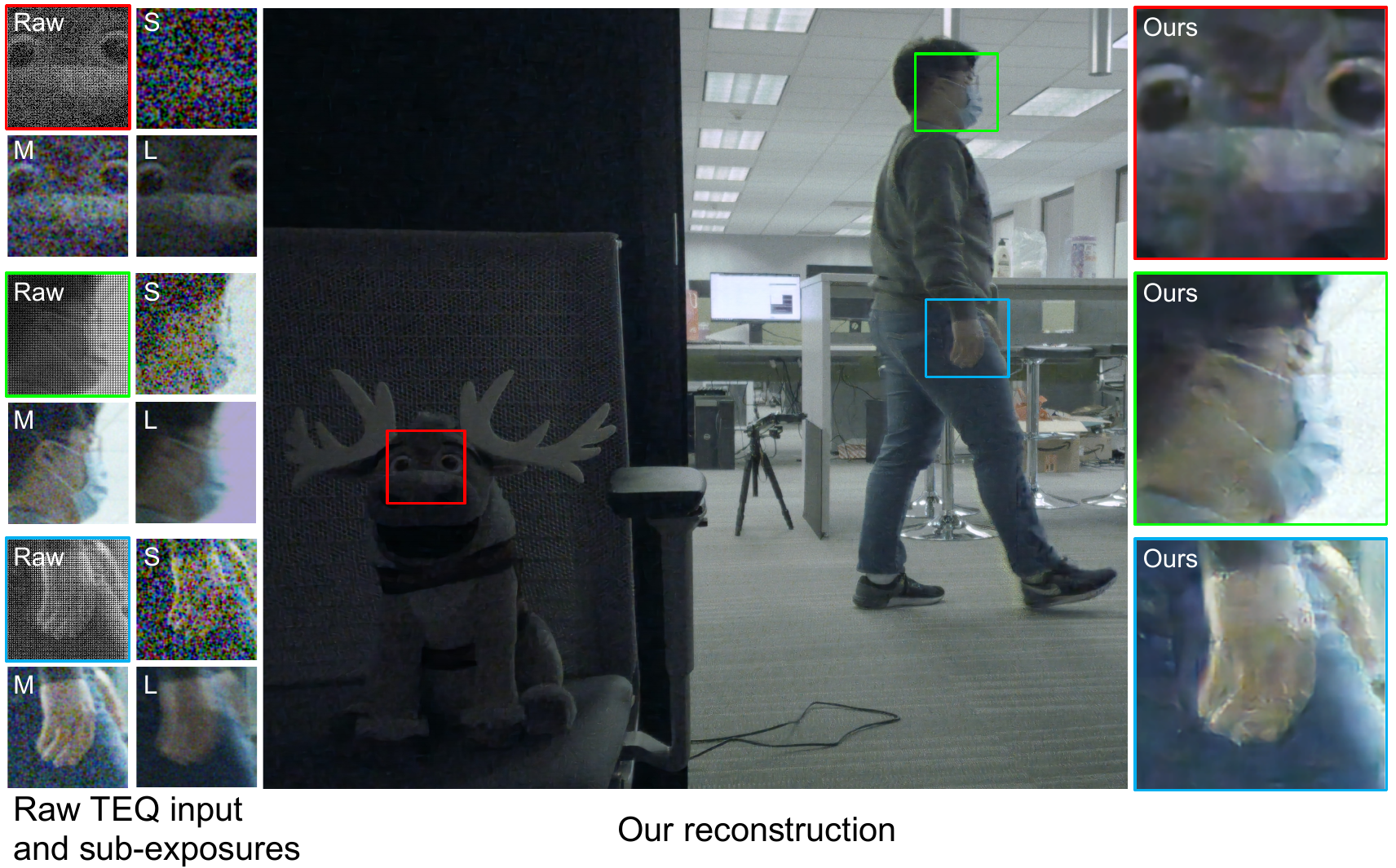}
\vspace{-2mm}
\caption{\NEW{Our HDR reconstruction on a real sensor (Sony IMX708) output. Together with our reconstruction, the raw input and its sub-exposures are shown on the left. Note that the motion of the person induced the severe blur in L.}
}
\label{figure:result:real}
\vspace{-3.5mm}
\end{figure}

Figure~\ref{result:comp_state_of_the_art} shows the comparison on challenging scenes with significant darkness/saturation and complex motions. The TVE methods of Kalantari et al.~\cite{kalantari2019deephdrv} in (a) and that of Yan et al.~\cite{yan2019attention} in (b) suffer from ghosting artifacts for fast-moving objects. The ghosting is more severe when the reference frame is the long exposure frame among the alternating exposures, since it contains more motion blur. Also, Kalantari's method shows noise shown in the second row of (a), because the model does not have any denoising module. The burst shot methods of Hasinoff et al.~\cite{hasinoff2016burst} in (c) and Tassano et al.~\cite{tassano2020fastdvdnet} in (d) are not able to recover the details in the dark area, since the burst shot methods focus on the motion with the expense of the severe quantization in the dark part. In contrast, our reconstruction of the tri-exposure quad-bayer in (e) outperforms other methods by avoiding the ghosting artifacts and the noise in the dark area.  \NEW{In the quantitative evaluation shown in Table~\ref{table:comp_state_of_the_art}, our method shows the best scores among state-of-the-art HDR video methods.}


\subsection{HDR Reconstruction on Real TEQ Inputs}
\NEW{We reconstructed HDR from real tri-exposure quad-bayer (TEQ) inputs. A Sony IMX$708$ sensor was used to capture challenging real-life scenes with a large dynamic range and complex motion. Two results are shown in Figure~\ref{fig:intro} and Figure~\ref{figure:result:real}. From the raw patches and the sub-exposure patches extracted from them, shown on the  left of the figure, saturation, severe noise, spatial artifact, and motion blur are observed in the inputs. For all of the magnified sample patches, our reconstruction is able to produce clean and sharp HDR images from severe corruption including noise and blur. More real results and videos will be available in the supplemental material.
}

\subsection{Model and Computation Complexity}
Our network consists of $2.41$ M parameters, which is smaller than $4.95$ M of Tassano's and $11.76$ M of Kalantari's. But Yan's model has the smallest number of parameters, $1.00$ M. However, our computation complexities are $14.56$ GFLOPS and $29.04$ GMADD (Giga Multiply-Add), and they are less than $25.45$ GFLOPS and $80.81$ GMADD of Yan's. Kalantari's method is computationally lightest by having $9.46$ GFLOPS and $22.00$ GMADD. Tassano's method requires $12.32$ GFLOPS and $24.59$ GMADD. Note that the complexity is measured for an image patch of $256\times 256$.

\section{Ablation Studies}

\begin{figure*}[t]
\centering
\includegraphics[width=0.99\linewidth]{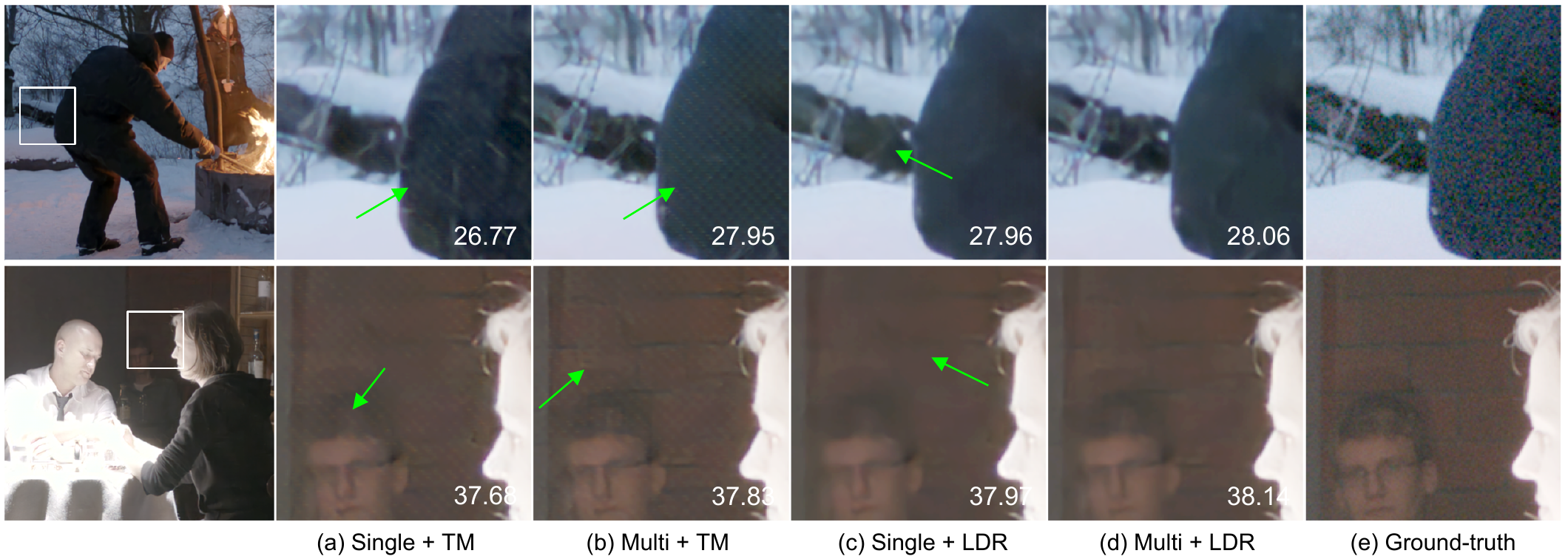}
\vspace{-1.0mm}
\caption{We studied the effect of multi-frame inputs and our LDR-reconstruction loss. The configuration of our proposed network, Multi + LDR, shows the best quality. Refer to Section~\ref{ablation:loss} for the details of each network.}
\label{result:ablation2}
\vspace{-3mm}
\end{figure*}

\subsection{Study on the Model Architecture}
\label{ablation:architecture}
We investigated the effect of the modules in our reconstruction network. We define a baseline reconstruction model that consists of an HDR fusion module and a temporal denoising module without any attention mechanisms. The HDR fusion module in the baseline does not perform the weight estimation, but it rather directly estimates an HDR feature. On the top of the baseline, we define a model called ATT by adding back the weight estimation and the attention mechanism to constitute exactly the same HDR fusion module and attention-based temporal denoising module as explained in Section~\ref{sec:ourmethod:hdrfusion} and~\ref{sec:ourmethod:temporaldenoising}. And NSR has a naive super-resolution module without the gate operation on top of ATT. In contrast, GSR+SSR network has a gated super-resolution  module that uses a stacked image of subsampled color and exposure pixels as the input of the super-resolution feature extractor. The stacked image is of the quarter size of the original quad-bayer image. Finally, GSR+OR network is our proposed one with the gated super-resolution utilizing the original quad-bayer image.

\begin{table}[t]
\centering
\resizebox{82mm}{!}{
\begin{tabular}{c | c c c c } 
 \hline
  & PSNR-$\mu$ & PuPSNR & HDR-VDP & HDR-VQM \\ 
 \hline\hline
 Baseline & $34.29$ & $38.69$ & $44.77$ & $0.8663$ \\ 
 ATT & $34.83$ & $39.53$ & $45.40$ & $0.8807$\\
 NSR & $35.32$ & $40.24$ & $45.91$ & $0.8924$ \\
 GSR + SSR & \textcolor{red}{$36.07$} &  \textcolor{blue}{$41.43$} & \textcolor{blue}{$45.97$} & \textcolor{blue}{$0.9215$} \\
 GSR + OR & \textcolor{blue}{$36.06$} &  \textcolor{red}{$41.64$} & \textcolor{red}{$46.16$} & \textcolor{red}{$0.9275$}  \\ 
 \hline
\end{tabular}}
\vspace{-1.0mm}

\caption{The network, GSR + OR, that has all of the proposed module shows the best scores. Refer to Section.~\ref{ablation:architecture} for the details of each network.}
\label{table:ablation1}
\vspace{-4.0mm}

\end{table}

\vspace{-3.0mm}
\paragraph{The Weight Estimation and the Attention}
By comparing (a) and (b) in Figure~\ref{result:ablataion1}, we observe ATT that has the weight estimation and the attention relatively less suffers from the artifacts in both dark and saturated region and helps to handle ghosting artifacts. The quantitative results in the first two rows of Table~\ref{table:ablation1} agree with the observation.

\vspace{-3.0mm}
\paragraph{The Explicit SR Module}
NSR that utilizes a naive explicit super-resolution module is clearly effective in improving the resolution and the blur problem as shown in Figure~\ref{result:ablataion1}(c). This is also validated by the third row in Table~\ref{table:ablation1}.

\vspace{-3.0mm}
\paragraph{The Gated SR Module}
Two networks, GSR+SSR and GSR+OR, that are equipped with the gated super-resolution module result in clear improvements over NSR with the naive super-resolution. In particular, for GSR+OR, we can further increase the resolution and the image quality by making use of the original quad-bayer input to extract super-resolution features. Compare (d) and (e) of Figure~\ref{result:ablataion1} and check out the quantitative results in Table~\ref{table:ablation1}.



\subsection{Study on the Inputs and the Loss functions}
\vspace{-1.0mm}
\label{ablation:loss}
We conducted an experiment to verify the effect of multi-frame inputs and the LDR-reconstruction loss proposed in Equation~\ref{eq:ldrreconloss}. Accordingly, we trained four different models shown in the first column of Table~\ref{table:ablation2}. The labels, Single and Multi, indicate the number of frames used as the inputs of the network. LDR stands out for the adoption of our LDR-reconstruction loss, and the naive tone-mapping loss function used in the previous works is indicated by TM.

\vspace{-3.0mm}
\paragraph{The Number of Input Frames}
The results of the multi-frame inputs in (b) and (d) of Figure~\ref{result:ablation2} recover the dark region better than those of the single-frame inputs in (a) and (c). In the quantitative evaluation, we could validate this effect as shown in the second and the fourth rows of Table~\ref{table:ablation2}. 

\vspace{-3.0mm}
\paragraph{The LDR-recon. Loss}
By comparing Figure~\ref{result:ablation2}(b) and (d), it is shown that the network trained with our LDR-reconstruction loss is more robust to the noise. Also, our proposed model labeled by Multi+LDR in Table~\ref{table:ablation2} shows the best scores of PSNR-$\mu$, PuPSNR, and HDR-VDP.

\begin{table}[t]
\centering
\resizebox{82mm}{!}{
\begin{tabular}{c | c c c c } 
 \hline
  & PSNR-$\mu$ & PuPSNR & HDR-VDP & HDR-VQM \\ 
 \hline\hline
 Single+TM & $35.89$ & $41.58$ & $46.06$ & $0.9107$ \\ 
 Multi+TM & \textcolor{blue}{$36.06$} & $41.64$ & $46.16$ & \textcolor{red}{$0.9275$} \\
 Single+LDR & $35.99$ & \textcolor{blue}{$42.01$} & \textcolor{blue}{$48.43$} & $0.9202$ \\
 Multi+LDR & \textcolor{red}{$36.42$} &  \textcolor{red}{$42.45$} & \textcolor{red}{$48.47$} & \textcolor{blue}{$0.9247$} \\
 \hline
\end{tabular}}
\vspace{-1.0mm}
\caption{The network taking multi-frame inputs trained with our LDR-reconstruction loss shows the best scores. Refer to Section~\ref{ablation:loss} for the details.}
\vspace{-4.0mm}

\label{table:ablation2}
\end{table}

\section{Conclusion}
\label{sec:discussion}
\noindent 
We proposed a novel network-based HDR video reconstruction for tri-exposure quad-bayer (TEQ) sensors. Our network is able to produce high-quality HDR video. The experiments showed that HDR video with TEQ is more optimal than the temporally varying exposure, the burst shot strategy, and the interlacing HDR, especially for scenes with large dynamic range and moving objects. Our ablation studies validated the functionality of modules in our network and the proposed LDR-reconstruction loss function.


\clearpage
{\small
\bibliographystyle{ieee_fullname}
\bibliography{hdrvideobib}
}

\end{document}